\documentclass[runningheads,a4paper]{llncs}
\usepackage{etex}
\reserveinserts{28}
\pdfoutput=1

\usepackage[linesnumbered,lined,boxed,commentsnumbered,ruled]{algorithm2e}

\SetAlFnt{\small}
\SetAlCapFnt{\small}
\SetAlCapNameFnt{\small}
\SetAlCapHSkip{0pt}
\IncMargin{-\parindent}
\usepackage{ textcomp }
\usepackage{ comment }
\usepackage[color=yellow!60,textsize=footnotesize,obeyDraft,draft]{todonotes}
\usepackage{colortbl}
\usepackage{booktabs}
\usepackage{capt-of}
\usepackage{makecell}
\usepackage{textcomp }
\usepackage{multirow}
\usepackage{listings}
\usepackage{graphicx}
\usepackage{colortbl}
\usepackage{booktabs}
\usepackage{amsmath}
\usepackage[font=footnotesize,labelfont=bf]{subcaption}
\usepackage[hidelinks]{hyperref}
\usepackage[capitalize]{cleveref}
\usepackage{tikz}
\usepackage{pgfplots}
\usepackage[para,online,flushleft]{threeparttable}
\usepackage{multirow}
\usepackage{wasysym}
\usepackage{amssymb}
\usepackage{enumitem}
\usetikzlibrary{arrows}
\usepackage{subcaption}
\usepackage{breakcites}
\usepackage{url}
\newcommand*\rot{\rotatebox{90}}
\newcounter{mnotei}
\newcolumntype{L}[1]{>{\raggedright\let\newline\\\arraybackslash\hspace{0pt}}m{#1}}
\newcolumntype{C}[1]{>{\centering\let\newline\\\arraybackslash\hspace{0pt}}m{#1}}
\newcolumntype{R}[1]{>{\raggedleft\let\newline\\\arraybackslash\hspace{0pt}}m{#1}}

\newcommand{\includegraphicsmaybe}[2]{
    \IfFileExists{#2}{\includegraphics[#1]{#2}}{
    \detokenize{File #2 is missing, maybe you need to run plots.py?}
}}

\begin{document}

\mainmatter

\title{Energy-Optimal Configurations for Single-Node HPC Applications}

\titlerunning{Energy-Optimal Configurations for Single-Node HPC Applications}

\author{Vitor~R.~G.~Silva\inst{1}, Alex Furtunato\inst{1}, Kyriakos Georgiou\inst{2}, Kerstin Eder\inst{2}, Samuel~Xavier-de-Souza\inst{1}}

\authorrunning{Vitor~R.~G.~Silva et al.}
\institute{Universidade Federal do Rio Grande do Norte, Brazil \and University of Bristol, UK}
\tocauthor{Authors' Instructions}
\maketitle

\makeatletter
\renewcommand\subsubsection{\@startsection{subsubsection}{3}{\z@}%
                       {-18\p@ \@plus -4\p@ \@minus -4\p@}%
                       {4\p@ \@plus 2\p@ \@minus 2\p@}%
                       {\normalfont\normalsize\bfseries\boldmath
                        \rightskip=\z@ \@plus 8em\pretolerance=10000 }}
\makeatother

\begin{abstract}
Energy efficiency is a growing concern for modern computing, especially for HPC due to operational costs and the environmental impact. We propose a methodology to find energy-optimal frequency and number of active cores to run single-node HPC applications using an application-agnostic power model of the architecture and an architecture-aware performance model of the application. We characterize the application performance using Support Vector Regression. The power consumption is estimated by modeling CMOS dynamic and static power without knowledge of the application. The energy-optimal configuration is estimated by minimizing the product of the power model and the performance model's outcomes. Results for four PARSEC applications with five different inputs show that the proposed approach used about 14$\times$ less energy when compared to the worst case of the default Linux DVFS governor. For the best case of the DVFS scheme, 23\% savings were observed, with an overall average of 6\% less energy.
\end{abstract}

\section{Introduction} \label{sec:introduction}
Processors are the main contributor to the power consumption of High Performance Computing (HPC) servers. They contribute between 20 and 40\% to the total server’s power draw \cite{Fan2007}. Google's servers showed that during peak utilization processors consumed about 57\% of the overall server’s power consumption \cite{Barroso2007}. Reducing processor power consumption is an effective approach to reduce the whole system's power consumption.  Therefore, modern processors incorporate several features for power management such as independent processing cores that can be disabled by the operating system \cite{Rotem2012}, clock gating techniques for reducing the dynamic power dissipation of synchronous circuits \cite{Srinivasan2015} and Dynamic Voltage and Frequency Scaling (DVFS) \cite{Mittal2014}.

DVFS has been demonstrated to be a very effective technique for reducing the power consumption of processors \cite{Hackenberg2015, Dzhagaryan2014, Hahnel2012, Basmadjian2012, Travers2015, Miyoshi2002, Anghel2011, Pietri2014}. The technique tries to optimize power consumption by adjusting the frequency according to the current load of the processor. Generally, the frequency scales with the intensity of the load and the voltage scales to the minimum value that enables the selected frequency. Among other aspects, DVFS helps reducing energy consumption because it allows memory-bounded programs to be executed more efficiently \cite{Spiga2006}. Nonetheless, aspects such as load variability may compromise the effectiveness of DVFS. Another important aspect that is typically not taken into account is the number of processing cores to be used by a parallel program. This choice is left to the user, which often is not trivial as shown in this paper.

We propose a methodology to find the operating frequency and number of active cores that minimize the total energy used to execute an HPC application on a single shared-memory HPC node.

The methodology uses an application-agnostic power model and an architecture-specific application characterization to model performance. The power model is based on the modeling of Complementary Metal-Oxide-Semiconductor (CMOS) logic in function of the operating frequency~\cite{Sarwar1997}. It models both the dynamic and static power. Besides operating frequency, the power model is also parametric to the number of active sockets and the number of active cores per socket. 

Performance is modeled by characterizing the application on the target architecture. The idea is to predict the performance of the application at any given configuration. The model takes as inputs the operating frequency, the number of active cores and the input size. The modeling is done using a supervised learning method for regression called Support Vector Regression (SVM)~\cite{Ventura2009, Smola2004}.

To find the optimal-energy configurations, the algorithm minimizes the product of outcomes of the power and performance models. This approach was validated on four PARSEC applications~\cite{Bienia2008} and compared to the \emph{Ondemand} governor, which is the default DVFS scheme for the Linux operating system. The results show that the proposed approach was able to find configurations that used about 14$\times$ less energy when compared to the worst case of the Ondemand governor. When compared to the best case of this DVFS scheme, i.e. when the user guesses the optimal number of cores to be used, the proposed approach was able to find configurations that used as much as 23\% less energy to execute the target application. The overall average energy saving reached 6\% for the proposed approach when compared to the best case and 790\% when compared to the worst case.

The rest of this paper is organized as follows. Section~\ref{sec:models} presents the proposed models for power, performance, and energy. The experimental setup and the fitting of the models are described in Section~\ref{sec:experimentalsetup}. In Section~\ref{sec:experimentsresults}, the results of applying the proposed approach to four PARSEC applications are presented. Related works are presented in Section~\ref{sec:relatedwork}. Finally, conclusions are drawn and future work is proposed in Section\ref{sec:conclusion}.

\section{Models} \label{sec:models}
In this Section, we present the proposed power and performance models that are used to estimate the minimum-energy consumption configuration.

\subsection{Power Model} \label{sec:powermodel}
Some of the main factors that contribute to the CPU power consumption are the dynamic power consumption, the short-circuit power consumption, and the power loss due to the current leakage of transistors, \cite{Rauber2014, Goel2016, Du2017, Gonzalez1997}. The complexity of the circuits of modern processors makes it very difficult to model their power consumption accurately. A viable approach for modeling the CPU's power draw is to model their building components, which are mainly made out of CMOS logic gates. Thus, modeling the power consumption for one logic gate and multiplying this by the total number of gates reduces the complexity of modeling the internal circuits but still provides the sufficient accuracy needed for making optimization decisions.

There are three main components of power dissipation in digital CMOS circuits,
\begin{equation}
P_{total}=P_{static}+P_{leak}+P_{dynamic} \label{eq_totalPower}
\end{equation}
namely, static power $P_{static}$, dynamic power $P_{dynamic}$, and leakage power $P_{leak}$.
According to~\cite{Sarwar1997, Butzen2007}, the dynamic power and leakage power behavior can be approximated by:
\begin{equation}
P_{dynamic}=CV^2f \label{eq_powerdyn},
\end{equation}
and
\begin{equation}
P_{leak} \propto V \label{eq_poweleak},
\end{equation}
where $C$ is the CMOS capacitance, $V$ the voltage applied to the circuit and $f$ the switching frequency.

Another common approximation is to expect a linear relationship between the voltage and the applied frequency~\cite{Usman2013}:
\begin{equation}
f \propto V \label{eq_fapoxV}
\end{equation}

Thus, the proposed model for one processing core of a multi-core processor is derived by using (\ref{eq_powerdyn}), (\ref{eq_poweleak}) and (\ref{eq_fapoxV}) to rewrite (\ref{eq_totalPower}) as follows:

\begin{equation}
P_{total}(f)= c_1f^3+c_2f+c_3 \label{eq_fitting},
\end{equation}

where $c_1$, $c_2$, and $c_3$ are the model's parameters.

When we include the number of active cores $p$, the estimation of the power consumption of the whole processor becomes:
\begin{equation}
P_{total}(f,p)= p(c_1f^3+c_2f)+c_3 \label{eq_power_core}.
\end{equation}

For systems that have more than one processor sockets, the power cost of enabling each socket can be considered. Adding the number of sockets $s$ to the equation gives the final version of the power model used in this work:
\begin{equation}
P_{total}(f,p,s)= p(c_1f^3+c_2f)+c_3+c_4s \label{eq_power_final},
\end{equation}
with $c_4$ being the model parameter for the number of sockets.

\subsection{Performance Model} \label{sec:performancemodel}
The performance model aims to estimate the application's execution time for a given target architecture based on a given operating frequency, number of active cores and input size. 

The performance was modeled by sampling the execution time of the application for several combinations of discrete values of frequency, number of active cores and input size. The samples where used as a training set for a Support Vector Regression (SVR); a version of the Support Vector Machine (SVM) algorithm for regression proposed in~\cite{Drucker1997}.

Training the SVR means minimize the weights $w$ subject to:

\[ \begin{cases} 
      y_i-\langle w,x_i\rangle-b \leq \varepsilon \\
      \langle w,x_i\rangle+b-y_i \leq \varepsilon \\
   \end{cases}
\]

In our model $x_i$ is a vector with the frequency, number of active cores and input size, $y_i$ is the execution time measured. $\langle w,x_i\rangle+b-y_i$ is the predicted output time and $\varepsilon$ is a free parameter that serves as a threshold.

\subsection{Energy Model} \label{sec:energymodel}
By combining outcome of the power model described in~\cref{sec:powermodel} and the SVR characterization of the application performance described in \cref{sec:performancemodel}, we can estimate the total energy used by the application as follows:
\begin{equation}
E(f,p,s,N)=P(f,p,s)\times{\rm SVR}(f,p,N) \label{eq_energy},
\end{equation}
where $P(f,p,s)$ is the total power modeled by~(\cref{eq_totalPower}), ${\rm SVM}(f,p,N)$ is the execution time estimated by the SVR characterization of the application, $f$ is the frequency, $p$ is the number of active cores, $s$ is the number of sockets, and $N$ is the input size. 

With (\cref{eq_energy}), it is possible to calculate energy consumption estimations for every possible configuration. Then, the configuration that minimizes energy consumption for a given input can be selected. It is also possible to apply constraints on the execution time, frequency, and the number of active cores although this is not considered in this work.

\section{Experimental Setup} \label{sec:experimentalsetup}

In the following subsections we present the software and hardware experimental setup used to validate the proposed approach.

\subsection{Case-Study Applications} \label{sec:casestudyapplication}

Four applications from the PARSEC parallel benchmark suite, version 3.0~\cite{Bienia2008}, were used as case-studies. 
This suite focuses on emerging workloads and was designed to be representative of the next generation shared-memory programs for chip-multiprocessors. The four applications used in this work were chosen for being relatively straightforward to devise smaller input sizes from the standard native inputs. These are: Fuidanimate, Raytrace, Swaptions, and Blackscholes. A short description of each one follows.

\subsubsection{Blackscholes} calculates the prices for a portfolio of European options analytically using the Black-Scholes partial differential equation. There is no closed-form expression for the Black-Scholes equation and as such it must be computed numerically. The program's inputs are the number of threads, the input file containing the options data, and the output file name.

\subsubsection{Fuidanimate} uses an extension of the Smoothed Particle Hydrodynamics (SPH) method to simulate an incompressible fluid for interactive animation purposes. The inputs are the number of threads, the number of frames, and an input file with information of all fluid particles and his proprieties.

\subsubsection{Raytrace} is a version of the raytracing method that is typically employed by real-time animations such as the ones used in computer games. It is optimized for speed rather than realism. The computational complexity of the algorithm depends on the resolution of the output image and the scene. The inputs used on this applications was the number of threads, the number of frames, a 3D object and the display resolution.

\subsubsection{Swaptions} Uses the Heath-Jarrow-Morton (HJM) framework to price a portfolio of swaptions. Swaptions employs Monte Carlo (MC) simulation to compute the prices. The input to this program are the number of threads, number of swaptions and the number of trials.

\subsection{Case-Study Architecture} \label{sec:casestudyarchitecture}
In the experiments performed in this work, we used compute nodes that consists of two Intel Xeon E5-2698 v3 processors with sixteen cores each and two hardware threads for each core. The maximum non-turbo frequency is 2.3GHz, and the total physical memory of the node is 128GB (8$\times$16GB). Turbo frequency and hardware multi-threading were disabled during all experiments. The operating system used is Linux CentOS 6.5, kernel 2.6.32.

The Linux kernel has many drivers available developed by the CPU manufacturers and the community \cite{Brown2005}. The default driver is the "acpi-cpufreq'' that uses policies implemented by so-called governors that dynamically decide the frequency values. Some of the governors available are Performance, Powersave, Ondemand, Conservative and Userspace. Performance and Powersave are static, and they set the frequency to the maximum and minimum allowed values, respectively. Ondemand and Conservative implement algorithms to estimate the CPU required capacity and adjust the processor frequency accordingly. Finally, Userspace allows the user to specify the frequency.

In this work, changing the frequency of the cores was done using the Linux "acpi-cpufreq'' driver. The number of active cores was changed by modifying the appropriate Linux virtual files. Both changes require root privileges. In practice, this approach can be brought into production by allowing the resource manager to perform this changes for the user using pre- and post-scripts for job submissions with energy consumption requirements.

\subsection{Fitting the Power Model} \label{sec:powerfitting}
To fit the power-model equation, the CPU was stressed up to 100\% and power information was acquired from the Intelligent Platform Management Interface (IPMI) sensors with a sampling rate of about one sample per second. IPMI provides information about variables and resources such as the system's temperature, voltage, fans and power supplies; using independent sensors attach to the hardware.

The power was collected for all combinations of frequency --- starting from 1.2~GHz and increasing by 100 MHz each time until 2.2~GHz is reached, and possible numbers of active cores --- from 1 to 32. Between each test the CPU was left idle until it cooled down to avoid interference on the next test.

\begin{figure}
\centerline{\includegraphics[width=0.8\linewidth]{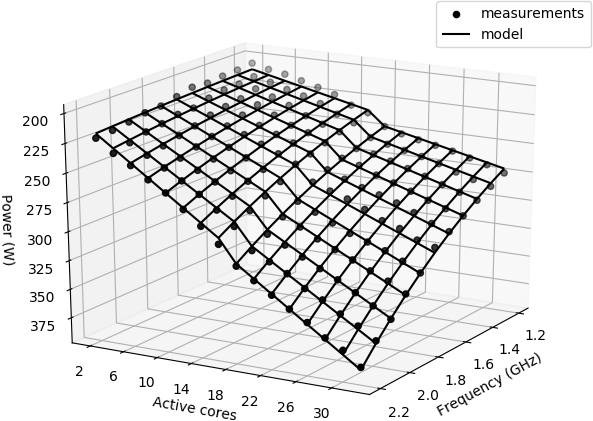}}
\caption{Power model fitting. The dots represent real power measurements and the solid lines represents the modeled power.}
\label{fig:fitting}
\end{figure}

The coefficients of (\ref{eq_power_final}), $c_1$, $c_2$, $c_3$ and $c_4$, were found by performing multi-linear regression on the data collected. The retrieved fitting can be seen on Fig.~\ref{fig:fitting}.

The equation for estimating the power in the target architecture turned to be:
\begin{equation} 
P_{total}(f,p,s)=p(0.29f^3+0.97f)+198.59+9.18s, \label{eq:fittedpower}
\end{equation}
where the unit for frequency is GHz.

To validate this model was calculated the absolute percentage error, i.e. the mean of the perceptual error on each point. This metric was chosen because of the significant difference between the smallest and the biggest values and it is calculated as follows:

\begin{equation}
\sum_{i}^{\rm \# samples}{\frac{|y_i-y_{\rm model}|}{y_i}}.
\label{mpe}
\end{equation}

The resulting absolute percentage error was 0.75\% and the root-mean squared error was 2.38W.

\subsection{Performance Characterization} \label{sec:performancecharacterization}

To characterize an application, we ran it for all different numbers of active cores in the range of $1<=p<=32$, for all the frequencies in the range of $1.2<=f<=2.2$ using 100MHz steps, and for 5 different input sizes.

The input sizes were chosen in such a way that the average execution time was in the order of minutes. The sampled power information, on every second, was used to calculate the real energy usage. The total time to complete the characterization varied between one and two days, depending on the application.

The SVR model was built using the collected data. A grid search was used to tune the model parameters. In this case, a Radial Base Function (RBF) kernel and the penalty for the wrong term of $10\times10^3$ and gamma 0.5 \cite{scikit-learn}. To train the SVR, the data collected was divided into two parts, 90\% for training and 10\% to test the accuracy.

The model was validated also using a cross-validation $k$-fold with $k$ equal to 10, using the Mean Absolute Error (MAE) and Percentage Absolute Error (PAE) as metrics. The average results of the cross validation can be seen in Table~\ref{tab:svr_evaluation}.

\begin{table}
\centering
\caption{Performance-Model's Cross validation Errors}
\label{tab:svr_evaluation}
\begin{tabular}{|l|l|l|l|l|}
\hline
\hline
Application  & MAE & PAE \\ \hline
Blackscholes & 2.01  & 4.6\% \\ \hline
Fluidanimate & 6.65  & 1.89\% \\ \hline
Raytrace  & 3.77  & 0.87\% \\ \hline
Swaptions & 2.29  & 2.56\% \\ \hline
\end{tabular}
\end{table}

\begin{figure}
	\centerline{\includegraphics[width=0.8\linewidth]{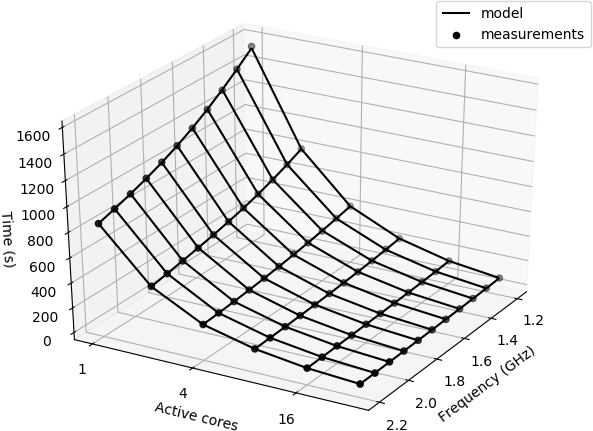}}
    \caption{Fluidanimate's performance model. The dots represent real performance measurements and the solid lines represent the modeled performance for various numbers of active cores and frequencies when running for input size 3.}
	\label{fig:svr_time2}
\end{figure}

\begin{figure}
	\centerline{\includegraphics[width=0.8\linewidth]{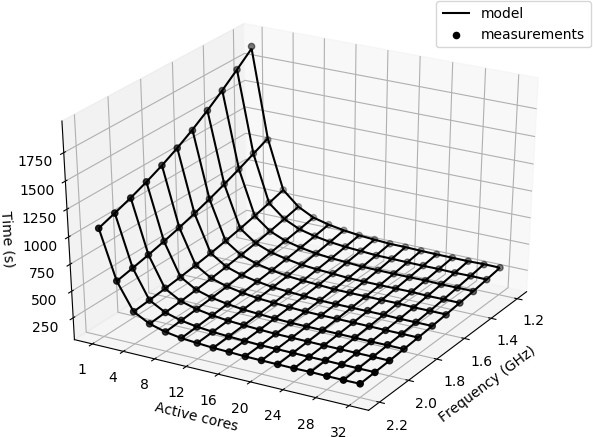}}
	\caption{Raytrace's performance model. The dots represent real performance measurements and the solid lines represent the modeled performance for various numbers of active cores and frequencies when running for input size 3.}
	\label{fig:svr_time3}
\end{figure}

\begin{figure}
	\centerline{\includegraphics[width=0.8\linewidth]{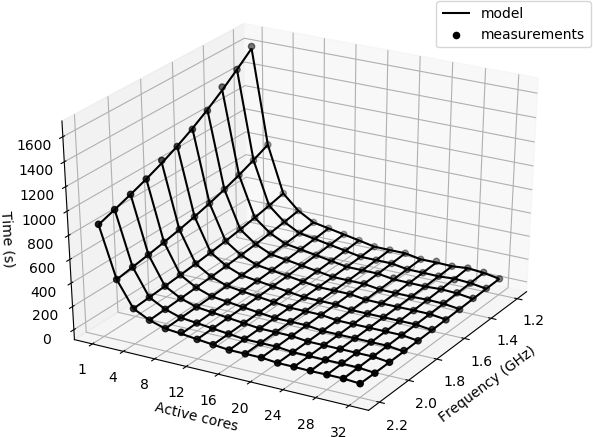}}
	\caption{Swaptions' performance model. The dots represent real performance measurements and the solid lines represent the modeled performance for various numbers of active cores and frequencies when running for input size 3.}
	\label{fig:svr_time4}
\end{figure}

\begin{figure}
	\centerline{\includegraphics[width=0.8\linewidth]{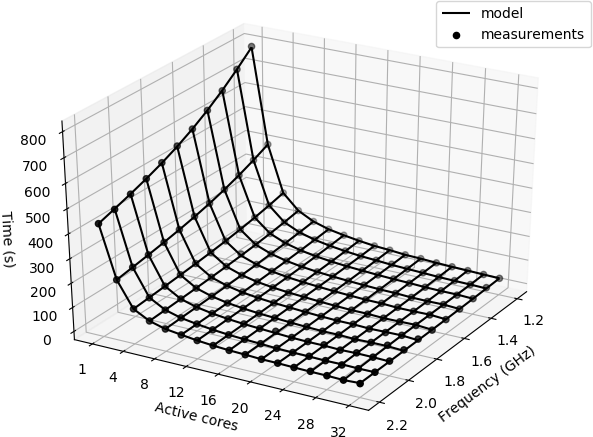}}
    \caption{Blackscholes' performance model. The dots represent real performance measurements and the solid lines represent the modeled performance for various numbers of active cores and frequencies when running for input size 3.}
	\label{fig:svr_time1}
\end{figure}

The results of the characterization can be seen in Figs.~\ref{fig:svr_time2},~\ref{fig:svr_time3},~\ref{fig:svr_time4}, and ~\ref{fig:svr_time1}.

\section{Experimental Results} \label{sec:experimentsresults}
In this Section, we present results for the energy model that we introduced in Section~\ref{sec:models} based on the parameter fitting described in Sections~\ref{sec:powerfitting} and~\ref{sec:performancecharacterization}. First, we compare and comment the model in contrast with the actual energy measurements. Finally, we evaluate the effectiveness of the proposed approach by comparing it to the Linux default Ondemand DVFS governor.

\subsection{Measured versus Modeled Energy} \label{sec:measuredversusmodeledenergy}
The energy measurements were obtained by integrating the power measurements over the total execution time of the application.
The power measurements were made using the IPMI sensors with a sampling rate of about one sample per second.

\begin{figure}
	\centerline{\includegraphics[width=0.8\linewidth]{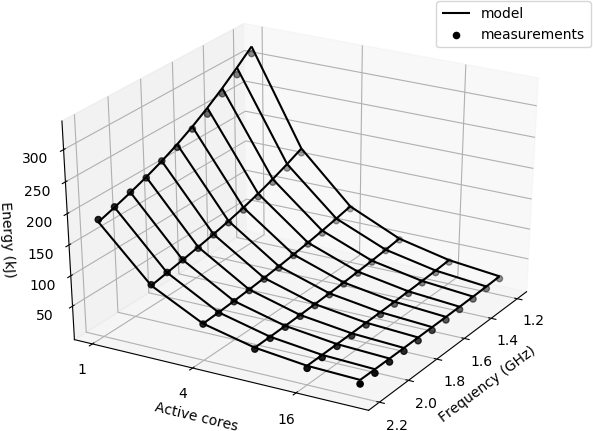}}
    \caption{Fluidanimate's energy measurements versus modeled energy consumption  varying the number of active cores and operating frequency, running with the input size 3.}
	\label{fig:fluid_s3}
\end{figure}

\begin{figure}
	\centerline{\includegraphics[width=0.8\linewidth]{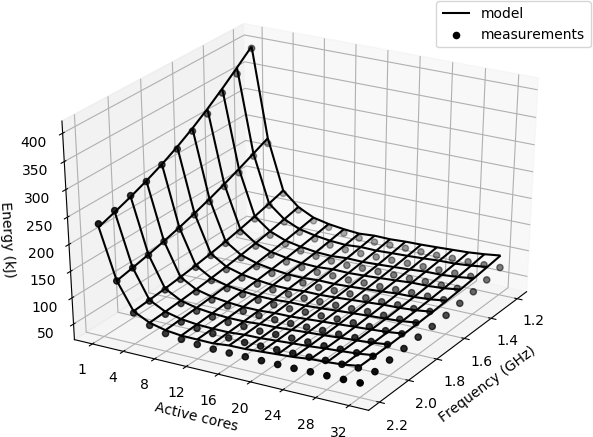}}
    \caption{Raytrace's energy measurements versus modeled energy consumption  varying the number of active cores and operating frequency, running with the input size 3.}
	\label{fig:rt_s3}
\end{figure}

\begin{figure}
	\centerline{\includegraphics[width=0.8\linewidth]{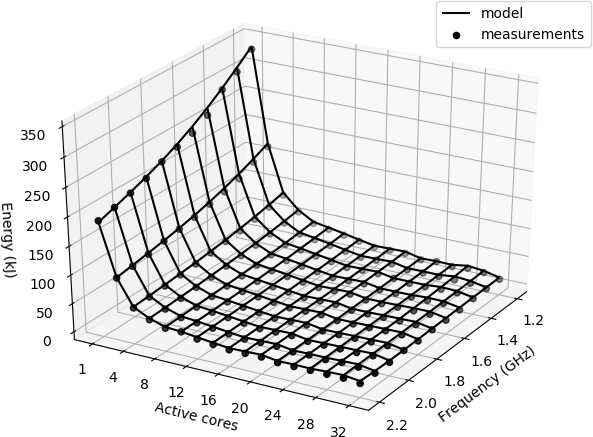}}
    \caption{Swaptions' energy measurements versus modeled energy consumption  varying the number of active cores and operating frequency, running with the input size 3.}
	\label{fig:swap_s3}
\end{figure}

\begin{figure}
	\centerline{\includegraphics[width=0.8\linewidth]{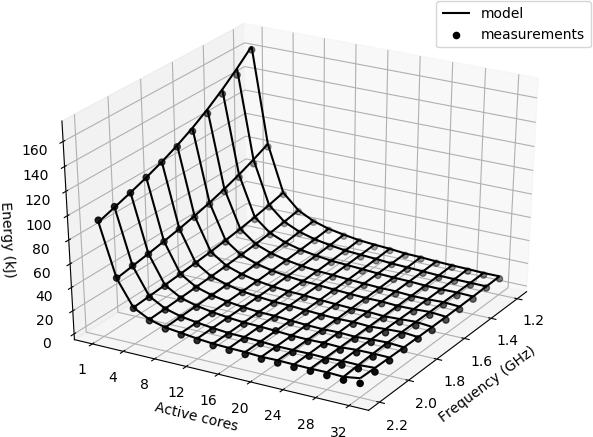}}
    \caption{Blackscholes' energy measurements versus modeled energy consumption varying the number of active cores and operating frequency, running with the input size 3.}
	\label{fig:black_s3}
\end{figure}

Figs. \ref{fig:fluid_s3}, \ref{fig:rt_s3}, \ref{fig:black_s3}, and~\ref{fig:swap_s3} plot the measured and modeled energy consumption for Blackscholes, Fuidanimate, Raytrace, and Swaptions, respectfully, for varying the number of active cores and operating frequency, running with the mid-size input.

In general, for the case-study applications and case-study architecture, the optimal-energy configurations tend to be the ones using the highest frequency, which characterizes a race-to-idle rather than a pace-to-idle optimal behavior~\cite{kim2015racing}. This can be explained by the large static power observed in the considered architecture, evidenced by the large $c_3$ parameter in (\ref{eq_power_final}) that was fitted in (\ref{eq:fittedpower}). With a large static power, using a pace-to-idle strategy, i.e. the use of frequencies lower than the maximum, is expected to be effective only if the sum of the leakage and the dynamic power parcels is larger than the static power parcel. 
Based on the fitted power model, this would never happen, i.e. the sum of leakage and dynamic power is always less than the static power,
\begin{equation*}
p(0.29f^3+0.97f)+9.18s < 198.59, 
\end{equation*}
even if we use the maximum number of cores, $p=32$ and $s=2$, and the maximum frequency, $f=2.2$.
Nevertheless, race-to-idle was not always the best strategy because energy scales with the execution time, which in turn scales inversely with the number of active cores and the operating frequency, and because power scales linearly with the number of cores, but exponentially with the frequency.

The optimal number of active cores depends on the parallel scalability of the application. The more scalable the application, the more cores it requires to minimize energy. A scalable application can increasingly exchange the speedup of more cores with lower frequencies in order to spend less energy. This is because of the linear relationship between power and number of cores and the exponential relationship between power and frequency.

\subsection{Proposed Approach versus Ondemand Linux Governor}
\label{sec:proposedapproach}
We have compared the energy consumption of the four case-study applications using the energy-optimal configurations provided by the proposed approach to the energy consumption resulted by use of the Linux default DVFS governor, Ondemand. Since the governor does not choose the number of active cores, we executed each application using 1, 2, 4, 8,$\cdots$, 28, 30, and 32 cores, accounting for the best and the worst cases of energy consumption. Tables \ref{tab:fluidfreq}, \ref{tab:raytracefreq}, \ref{tab:swapfreq} and \ref{tab:blackfreq} present these results for Fuidanimate, Raytrace, Swaptions, and Blackscholes, respectively. \begin{table}
\caption{Fluidanimated Minimal energy}
\label{tab:fluidfreq}
\resizebox{\columnwidth}{!}{%
\begin{tabular}{l|l|l|l|l|l|l|ll}

\multicolumn{1}{l|}{\rot{Input}} & \rot{\begin{tabular}[c]{@{}l@{}}Mean Freq. \\ in GHz \\ (\#Cores) \end{tabular}}  & \rot{Energy in KJ}   &  \rot{\begin{tabular}[c]{@{}l@{}}Mean Freq. \\ in GHz \\ (\#Cores) \end{tabular}} & \rot{Energy in KJ}  & \rot{\begin{tabular}[c]{@{}l@{}} Freq. \\ in GHz \\ (\#Cores) \end{tabular}} & \rot{Energy in KJ}        & \multicolumn{1}{l|}{\rot{Min. Save(\%)}} & \multicolumn{1}{l}{\rot{Max. Save(\%)}} \\ \hline

\multicolumn{1}{l|}{1}     & 1.85 (32)        & 4.85            & 2.29 (1)         & 32.38           & 2.0 (32)  & 4.15        & \multicolumn{1}{l|}{16.90}       & \multicolumn{1}{l}{680.31}      \\ \hline
\multicolumn{1}{l|}{2}     & 1.88  (32)       & 9.35            & 2.29 (1)         & 66.77           & 2.0 (32)  & 7.89        & \multicolumn{1}{l|}{18.60}       & \multicolumn{1}{l}{746.54}      \\ \hline
\multicolumn{1}{l|}{3}     & 1.89  (32)       & 18.82           & 2.30 (1)         & 135.00          & 2.0 (32)  & 16.98       & \multicolumn{1}{l|}{10.86}       & \multicolumn{1}{l}{695.04}      \\ \hline
\multicolumn{1}{l|}{4}     & 2.08  (32)       & 37.80           & 2.30 (1)         & 272.55          & 2.1 (32)  & 33.20       & \multicolumn{1}{l|}{13.84}       & \multicolumn{1}{l}{720.82}      \\ \hline
\multicolumn{1}{l|}{5}     & 2.00  (32)       & 76.28           & 2.30 (1)         & 546.84          & 2.2 (32)  & 66.83       & \multicolumn{1}{l|}{14.14}       & \multicolumn{1}{l}{718.24} \\ \hline

& \multicolumn{2}{l|}{Ondemand Min.} & \multicolumn{2}{l|}{Ondemand Max.} & \multicolumn{2}{l|}{Proposed} &                             &                             \\ 

\end{tabular}
}
\end{table}
\begin{table}
\caption{Raytrace Minimal energy}
\label{tab:raytracefreq}

\resizebox{\columnwidth}{!}{%
\begin{tabular}{l|l|l|l|l|l|l|ll}

\multicolumn{1}{l|}{\rot{Input}} & \rot{\begin{tabular}[c]{@{}l@{}}Mean Freq. \\ in GHz \\ (\#Cores) \end{tabular}}  & \rot{Energy in KJ}   &  \rot{\begin{tabular}[c]{@{}l@{}}Mean Freq. \\ in GHz \\ (\#Cores) \end{tabular}} & \rot{Energy in KJ}  & \rot{\begin{tabular}[c]{@{}l@{}} Freq. \\ in GHz \\ (\#Cores) \end{tabular}} & \rot{Energy in KJ}        & \multicolumn{1}{l}{\rot{Save Min.(\%)}} & \multicolumn{1}{|l}{\rot{Save Max.(\%)}} \\ \hline

\multicolumn{1}{l|}{1}     & 1.30 (4)         & 38.56           & 2.29 (1)         & 60.29           & 2.2 (6)   & 37.92       & \multicolumn{1}{l|}{1,70}        & \multicolumn{1}{l}{59.01}       \\ \hline
\multicolumn{1}{l|}{2}     & 1.32 (8)         & 43.59           & 2.30 (1)         & 98.11           & 2.2 (10)  & 39.93       & \multicolumn{1}{l|}{9.16}        & \multicolumn{1}{l}{145.68}      \\ \hline
\multicolumn{1}{l|}{3}     & 1.65 (16)        & 49.40           & 2.30 (1)         & 168.82          & 2.2 (14)  & 45.77       & \multicolumn{1}{l|}{7.94}        & \multicolumn{1}{l}{268.84}      \\ \hline
\multicolumn{1}{l|}{4}     & 1.62 (32)        & 55.61           & 2.30 (1)         & 299.83          & 2.2 (22)  & 52.99       & \multicolumn{1}{l|}{4.94}        & \multicolumn{1}{l}{465.83}      \\ \hline
\multicolumn{1}{l|}{5}     & 1.77 (32)        & 69.33           & 2.30 (1)         & 520.34          & 2.2 (26)  & 67.28       & \multicolumn{1}{l|}{3.05}        & \multicolumn{1}{l}{673.39}      \\ \hline
& \multicolumn{2}{l|}{Ondemand Min.} & \multicolumn{2}{l|}{Ondemand Max.} & \multicolumn{2}{l|}{Proposed} &                             &                             \\ 
\end{tabular}
}
\end{table}
\begin{table}
\caption{Swaptions Minimal energy}
\label{tab:swapfreq}
\resizebox{\columnwidth}{!}{%
\begin{tabular}{l|l|l|l|l|l|l|ll}

\multicolumn{1}{l|}{\rot{Input}} & \rot{\begin{tabular}[c]{@{}l@{}}Mean Freq. \\ in GHz \\ (\#Cores) \end{tabular}}  & \rot{Energy in KJ}   &  \rot{\begin{tabular}[c]{@{}l@{}}Mean Freq. \\ in GHz \\ (\#Cores) \end{tabular}} & \rot{Energy in KJ}  & \rot{\begin{tabular}[c]{@{}l@{}} Freq. \\ in GHz \\ (\#Cores) \end{tabular}} & \rot{Energy in KJ}        & \multicolumn{1}{l|}{\rot{Min. Save(\%)}} & \multicolumn{1}{l}{\rot{Max. Save(\%)}} \\ \hline

\multicolumn{1}{l|}{1}     & 2.15 (32)        & 5.88            & 2.29 (1)         & 80.08           & 2.2 (32)  & 5.73        & \multicolumn{1}{l|}{2.57}        & \multicolumn{1}{l}{1297.82}    \\ \hline
\multicolumn{1}{l|}{2}     & 2.00 (32)        & 9,21            & 2.30 (1)         & 106.84          & 2.2 (32)  & 7,81        & \multicolumn{1}{l|}{17.90}       & \multicolumn{1}{l}{1267.59}    \\ \hline
\multicolumn{1}{l|}{3}     & 2.22 (32)        & 10.37           & 2.30 (1)         & 133.41          & 2.0 (32)  & 9.90        & \multicolumn{1}{l|}{4.70}        & \multicolumn{1}{l}{1247.58}    \\ \hline
\multicolumn{1}{l|}{4}     & 2.02 (32)        & 14.29           & 2.30 (1)         & 160.34          & 2.0 (32)  & 12.33       & \multicolumn{1}{l|}{15.95}       & \multicolumn{1}{l}{1200.85}    \\ \hline
\multicolumn{1}{l|}{5}     & 2.08 (32)        & 15.82           & 2.30 (1)         & 186.39          & 1.9 (32)  & 14.45       & \multicolumn{1}{l|}{9.50}        & \multicolumn{1}{l}{1190.15}    \\ \hline

& \multicolumn{2}{l|}{Ondemand Min.} & \multicolumn{2}{l|}{Ondemand Max.} & \multicolumn{2}{l|}{Proposed} &                             &                             \\ 

\end{tabular}
}
\end{table}
\begin{table}
\caption{Balckschoels Minimal energy}
\label{tab:blackfreq}
\resizebox{\columnwidth}{!}{%
\begin{tabular}{l|l|l|l|l|l|l|ll}

\multicolumn{1}{l|}{\rot{Input}} & \rot{\begin{tabular}[c]{@{}l@{}}Mean Freq. \\ in GHz \\ (\#Cores) \end{tabular}}  & \rot{Energy in KJ}   &  \rot{\begin{tabular}[c]{@{}l@{}}Mean Freq. \\ in GHz \\ (\#Cores) \end{tabular}} & \rot{Energy in KJ}  & \rot{\begin{tabular}[c]{@{}l@{}} Freq. \\ in GHz \\ (\#Cores) \end{tabular}} & \rot{Energy in KJ}        & \multicolumn{1}{l|}{\rot{Min. Save (\%)}} & \multicolumn{1}{l}{\rot{Max. Save(\%)}} \\ \hline

\multicolumn{1}{l|}{1}     & 1.57 (32)        & 1.36            & 2.27 (1)         & 16.35           & 2.2 (30)  & 1.69        & \multicolumn{1}{l|}{-19.32}      & \multicolumn{1}{l}{869.00}      \\ \hline
\multicolumn{1}{l|}{2}     & 2.09 (32)        & 2.93            & 2.24  (1)        & 33.16           & 1.8 (32)  & 3.36        & \multicolumn{1}{l|}{-12.78}      & \multicolumn{1}{l}{887.93}      \\ \hline
\multicolumn{1}{l|}{3}     & 1.82 (32)        & 8.08            & 2.23  (1)        & 65.97           & 2.2 (30)  & 6.55        & \multicolumn{1}{l|}{23.31}       & \multicolumn{1}{l}{907.02}      \\ \hline
\multicolumn{1}{l|}{4}     & 2.01 (32)        & 12.59           & 2.14  (1)        & 131.85          & 2.2 (26)  & 13.64       & \multicolumn{1}{l|}{-7.66}       & \multicolumn{1}{l}{866.97}      \\ \hline
\multicolumn{1}{l|}{5}     & 1.97 (32)        & 25.29           & 1.57  (1)        & 263.89          & 2.2 (28)  & 26.52       & \multicolumn{1}{l|}{-4.61}       & \multicolumn{1}{l}{895.23}      \\ \hline

& \multicolumn{2}{l|}{Ondemand Min.} & \multicolumn{2}{l|}{Ondemand Max.} & \multicolumn{2}{l|}{Proposed} &                             &                             \\

\end{tabular}
}
\end{table}

In most cases, the proposed approach obtained better results than the best cases of the Ondemand governor. For Blackscholes, the proposed approach was only better than the Ondemand best case for input number 3. On average, the proposed method was 6\% better than the best case of the Ondemand governor.

In all cases, the method proposed here outperformed the worst case of the Ondemand governor. On average, the difference in energy consumption was about 790\%, being 1298\% the maximum difference and 59\% the minimum. In general, the energy consumption of the DFVS scheme was larger for smaller numbers of cores. Nonetheless, it was not always the case that the best number of cores for this scheme was the maximum, i.e. 32 cores. Possibly, for architectures with larger number of cores, choosing the exact number the minimizes energy consumption would be less evident.

\begin{figure}[!htp]
    \centering
    \begin{subfigure}{\textwidth}
    \centering
    \begin{subfigure}{\textwidth}
    \centering
      \includegraphics[width=0.8\linewidth]{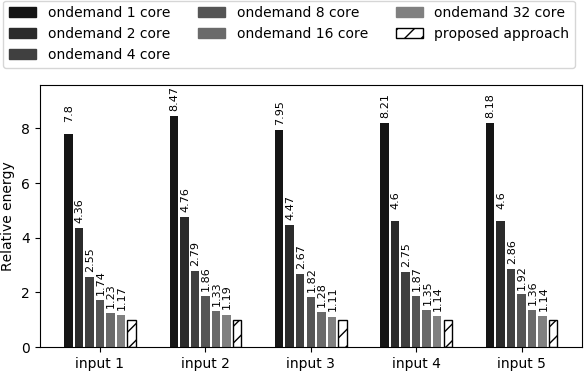}
    \end{subfigure} 
    \begin{subfigure}{\textwidth}
    \centering
        \includegraphics[width=0.8\linewidth]{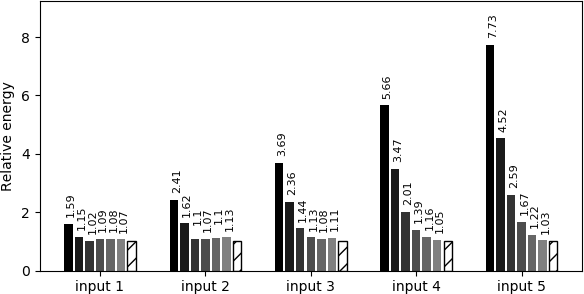}
    \end{subfigure}
    \end{subfigure}
    \begin{subfigure}{\textwidth}
    \centering
    \begin{subfigure}{\textwidth}
    \centering
      \includegraphics[clip=true,width=0.8\linewidth]{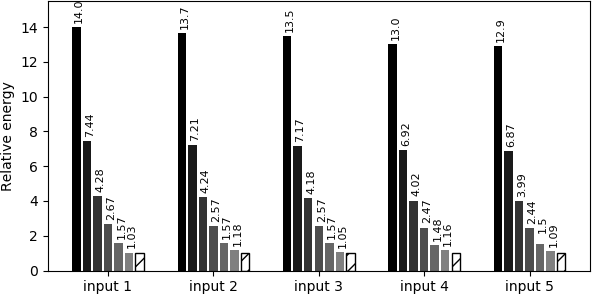}
    \end{subfigure} 
    \begin{subfigure}{\textwidth}
    \centering
        \includegraphics[clip=true,width=0.8\linewidth]{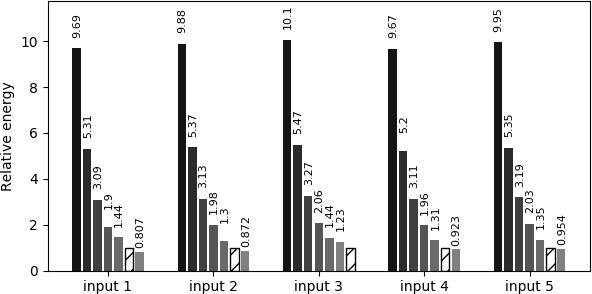}
    \end{subfigure}
    \end{subfigure}
    \caption{Energy consumption of the Ondemand governor for power-of-2 numbers of cores and the proposed approach. The values are relative to the energy of the proposed approach.}
	\label{fig:energy_results}
\end{figure}

Fig. \ref{fig:energy_results} shows the behavior of the energy consumption for all tested cases of the Ondemand governor and the proposed approach with values normalized to the energy consumption of the proposed approach.

\section{Related Work} \label{sec:relatedwork}
DVFS is the most common technique employed to obtain energy savings on multi-core systems. Thus, the technique has been extensively researched with the aim of providing strategies for selecting the optimal voltage and frequency for a specific application and architecture. In \cite{Anghel2011} the authors utilized two algorithms for scaling the frequency of the processors: a human-immune system inspired algorithm to monitor the server's power and performance states; and a fuzzy logic based algorithm for changing the server's performance state. \cite{Cochran2011} introduced a scaling method for determining the system's optimal operation points for the number of threads and DVFS settings.

In \cite{DaCosta2015}, an approach that considers instantaneous system activity states was proposed. In this case, the memory and network activity were used to generate a DVFS management setting.

Performance counters have also been used to perform effective DVFS. In \cite{Spiliopoulos2011}, the authors used a Continuous Adaptive DVFS based on a performance model of the processor. The model was based on sampling the hardware's performance counters at regular intervals to predict performance/energy workloads. Base on these predictions appropriate voltage, and frequency settings were selected.

In \cite{Georgiou:2017}, the authors used an energy model for a multi-threaded, multi-core embedded architecture and static resource analysis to statically evaluate the energy and timing savings of various DVFS configurations for the same program. Although, they were able to identify the most optimal configuration without the need of executing the program with each different configuration and measuring time and energy, there approach is quite limited as static analysis does not scale to less time predictable architectures and programs.

In this work, we introduce a power and a performance model to find energy-optimal operating frequency and number of active cores for applications running on specific multi-core platforms. Our approach does not use the DVFS manager to control the processor voltage and frequency settings. This new approach can obtain better results than DVFS strategies as was shown in \cref{sec:experimentsresults}. 

The success obtained from this approach is possibly due to the fact that the use previous knowledge of the application's performance on the target architecture can expose sufficiently relevant information, such as parallel speedups, that is harder to guess in runtime techniques based on DVFS.

The use of an application-agnostic power modeling for the target architecture helps to make the technique portable to other applications. That is, to estimate the energy-optimal frequency and number of active cores for a new application, only a performance characterization is needed.

\section{Conclusion and future work} \label{sec:conclusion}
In this paper, we propose a new approach to optimize the energy efficiency of single-node batch HPC applications. In contrast to existing scheduling algorithms, our technique utilizes the application's runtime profile, and a power model of the compute node to predict the optimal frequency and number of cores to be used. This proven effective in reducing the energy consumption of applications. 

Results from four parallel PARSEC applications running on an HPC node with two sixteen-core processors show that the novel approach outperforms the default Linux DVFS scheme on its best case with an average of 6\% energy savings. In its worst case, the savings were about 790\%, on average.

A weakness of the proposed technique is the need for information about the input size of the application before execution. A possible solution would be to use performance counters, present in all modern HPC processors, to guess the input size based on previously trained data.

Future work will improve the proposed energy model by taking into account more relevant information, such as the percentage of CPU utilization. This can enable the identification of different phases of the target program and thus, it will enable more fine-grained changes of the frequency and, perhaps, the number of active cores, to further improve the results presented here.

\section*{Acknowledgments}

The work is supported by the European Union’s Horizon 2020 Research and Innovation Programme under Grant agreement No.: 779882, TeamPlay (Time, Energy and security Analysis for Multi/Many-core heterogeneous PLAtforms), and by the Royal Society Newton Advanced Fellowship Programme under Grant No.: NA160108.

\bibliographystyle{alphaurl}
\bibliography{references}

\end{document}